%% file: HIN-18-015_temp.tex
\pdfoutput=1
\documentclass[11pt,twoside,a4paper,cmspaper,final,collab]{cms-tdr}
\def\svnVersion{d5d62cc-D}\def\svnDate{2020/11/15}\def\cmsCernNoTag{CERN-EP-2019-031}\def\cmsCernDate{\today}\def\cmsMessage{"Published in Physical Review C as \href{http://dx.doi.org/10.1103/PhysRevC.103.014902}{\doi{10.1103/PhysRevC.103.014902}.}"}
\begin{document}\cmsNoteHeader{HIN-18-015}

\newlength\cmsFigWidth
\ifthenelse{\boolean{cms@external}}{\setlength\cmsFigWidth{0.49\textwidth}}{\setlength\cmsFigWidth{0.65\textwidth}}
\ifthenelse{\boolean{cms@external}}{\providecommand{\cmsLeft}{upper\xspace}}{\providecommand{\cmsLeft}{left\xspace}}
\ifthenelse{\boolean{cms@external}}{\providecommand{\cmsRight}{lower\xspace}}{\providecommand{\cmsRight}{right\xspace}}
\ifthenelse{\boolean{cms@external}}{\providecommand{\NA}{\ensuremath{\cdots}\xspace}}{\providecommand{\NA}{\ensuremath{\text{---}}\xspace}}

\newcommand{\pp}{\ensuremath{\Pp\Pp}\xspace}
\newcommand{\pPb}{\ensuremath{\Pp\text{Pb}}\xspace}
\newcommand{\pA}{\ensuremath{\Pp\text{A}}\xspace}
\newcommand{\PbPb}{\ensuremath{\text{PbPb}}\xspace}
\newcommand{\AonA}{\ensuremath{\text{AA}}\xspace}

\newcommand{\deta}{\ensuremath{\Delta\eta}}
\newcommand{\dphi}{\ensuremath{\Delta\phi}}
\newcommand{\noff}{\ensuremath{N_\mathrm{trk}^\mathrm{offline}}\xspace}

\newcommand{\ra}{\rangle}
\newcommand{\lan}{\langle}
\newcommand{\mean}[1]{\lan #1 \ra}
\newcommand{\dmean}[1]{\lan\lan #1 \ra\ra}

\cmsNoteHeader{HIN-18-015}
\title{Correlations of azimuthal anisotropy Fourier harmonics with subevent cumulants in \texorpdfstring{\pPb collisions at $\sqrtsNN = 8.16\TeV$}{pPb collisions at sqrt(sNN) = 8.16 TeV}}

\date{\today}

\abstract{
Event-by-event long-range correlations of azimuthal anisotropy Fourier coefficients ($v_n$)
in 8.16\TeV \pPb data, collected by the CMS experiment at the LHC, are extracted using a subevent 
four-particle cumulant technique applied to very low multiplicity events.
Each combination of four charged particles are selected from either two, three, or four distinct subevent regions of
a pseudorapidity range from -2.4 to 2.4 of the CMS tracker, and with transverse momentum between 0.3 and 3.0\GeV.
Using the subevent cumulant technique, correlations between $v_n$ of different orders are measured as functions of
particle multiplicity and compared to the standard cumulant method without subevents over a wide event multiplicity range.
At high multiplicities, the $v_2$ and $v_3$ coefficients exhibit an anticorrelation; this behavior is 
observed consistently using various methods. The $v_2$ and $v_4$ correlation strength is found to depend 
on the number of subevents used in the calculation. As the event multiplicity decreases, the results from different 
subevent methods diverge because of different contributions of non-collective or few-particle correlations.
Correlations extracted with the four-subevent method exhibit a tendency to diminish monotonically toward the 
lowest multiplicity region (about 20 charged tracks) investigated. 
These findings extend previous studies to a significantly lower event multiplicity range 
and establish the evidence for the onset of long-range
collective multiparticle correlations in small system collisions.
}

\hypersetup{
pdfauthor={CMS Collaboration},
pdftitle={Correlations of azimuthal anisotropy Fourier harmonics in pPb collisions at sqrt(sNN) = 8.16 TeV},
pdfsubject={CMS},
pdfkeywords={CMS, heavy ion, ridge, high multiplicity, small system, flow}}

\maketitle

\section{Introduction}
\label{sec:intro}

In high-energy ultrarelativistic nucleus-nucleus (\AonA) collisions, a dense and hot state of matter called the quark gluon plasma (QGP) is produced~\cite{Shuryak:2004cy,Busza:2018rrf}.
Studies of multiparticle correlations provide important insights into the underlying mechanism of particle production in this strongly coupled, non-perturbative regime.
A key feature of such multiparticle correlations in \AonA collisions is a pronounced structure on the near side relative azimuthal angle ($\abs{\dphi}\approx0$)
that extends over a large range in relative pseudorapidity ($\abs{\deta}$ up to 4 units or more).
This feature, known as the ``ridge'', has been found over a wide range of center-of-mass energies and system sizes in \AonA collisions at both the BNL RHIC~\cite{Abelev:2009af,Alver:2008gk,Alver:2009id,Abelev:2009jv} and the CERN LHC~\cite{Chatrchyan:2011eka,Chatrchyan:2012wg,Aamodt:2011by,ATLAS:2012at,CMS:2013bza}. It is interpreted as arising primarily from the initial anisotropic geometry and its
fluctuations coupled with the collective hydrodynamic flow of a strongly interacting, expanding medium~\cite{Ollitrault:1992bk,Alver:2010gr}.
The azimuthal correlations of emitted particle pairs are typically characterized by their Fourier components as:
\begin{linenomath}
\begin{equation}
\frac{\rd{}N^\text{pair}}{\rd\Delta\phi} \propto 1 + \sum_{n} 2V_{n\Delta} \cos (n\Delta\phi),
\label{partdistcorr}
\end{equation}
\end{linenomath}
\noindent where $V_{n\Delta}$ are the two-particle Fourier coefficients. If factorization is assumed, $v_n = \sqrt{V_{n\Delta}}$ denote
the single-particle anisotropy harmonics~\cite{Voloshin:1994mz}. In particular, the second, third, and fourth Fourier components are known as elliptic ($v_2$), triangular
($v_3$), and quadrangular ($v_4$) flow, respectively~\cite{Alver:2010gr}.

In order to constrain the effects of the geometry and its fluctuations in the initial conditions, and the transport properties of the produced medium in \AonA collisions, new studies were
carried out looking at correlations between different orders of $v_n$ harmonics. In particular, event-by-event fluctuations of $v_n$
harmonic amplitudes in \PbPb collisions at the LHC were studied using the event shape engineering technique~\cite{Aad:2015lwa}, and the four-particle symmetric cumulant (SC)
method~\cite{ALICE:2016kpq, Bilandzic:2013kga}, where the SC method for two different harmonic orders $n$ and $m$ is defined as:
\ifthenelse{\boolean{cms@external}}{
\begin{linenomath}
\begin{equation}
\begin{aligned}
\text{SC}(n,m) & = \langle \langle \cos (n\phi_1 + m\phi_2 - n\phi_3 - m\phi_4) \rangle \rangle \\
               & - \langle \langle \cos (n\phi_1 - n\phi_2) \rangle \rangle \langle \langle \cos (m\phi_3 - m\phi_4) \rangle \rangle \\
               & = \langle v_n^2 v_m^2 \rangle - \langle v_n^2 \rangle \langle v_m^2 \rangle.
\end{aligned}
\label{scdef}
\end{equation}
\end{linenomath}
}{
\begin{linenomath}
\begin{equation}
\begin{aligned}
\text{SC}(n,m) & = \langle \langle \cos (n\phi_1 + m\phi_2 - n\phi_3 - m\phi_4) \rangle \rangle - \langle \langle \cos (n\phi_1 - n\phi_2) \rangle \rangle \langle \langle \cos (m\phi_3 - m\phi_4) \rangle \rangle, \\
               & = \langle v_n^2 v_m^2 \rangle - \langle v_n^2 \rangle \langle v_m^2 \rangle.
\end{aligned}
\label{scdef}
\end{equation}
\end{linenomath}
}
\noindent Here, the double angular brackets indicate that the averaging procedure is done first on all distinct 
particle quadruplets in an event, and then over all the events, by weighting each single event average with its number 
of quadruplets. Over the full range of impact parameters in \PbPb collisions, it was found that the $v_2$ harmonic
exhibits a negative event-by-event correlation with the $v_3$ harmonic, while the correlation is positive 
between the $v_2$ and $v_4$ harmonics. These correlations are shown
to be sensitive probes of initial-state fluctuations ($v_2$ \vs $v_3$) and medium transport coefficients
($v_2$ \vs $v_4$)~\cite{Alver:2010dn,Schenke:2010rr,Qiu:2011hf,Giacalone:2016afq,ALICE:2016kpq}.

In high-multiplicity \pp and \pA collisions, the ``ridge'' has been
observed~\cite{Khachatryan:2010gv,Khachatryan:2015lva,Aad:2015gqa,Aad:2014lta,Khachatryan:2014jra,Aaij:2015qcq,Adamczyk:2015xjc} and
detailed studies have highlighted its collective nature~\cite{Khachatryan:2015waa,Khachatryan:2016txc,Aaboud:2017acw,Aaboud:2018syf}. 
Event-by-event correlations among the $v_2$, $v_3$
and $v_4$ Fourier harmonics have also been measured for both systems using the SC method~\cite{Sirunyan:2017uyl}.
The correlation data reveal features similar to those observed in \PbPb collisions, where a negative correlation
is found between the $v_2$ and $v_3$ harmonics, while the correlation is positive between the $v_2$ and $v_4$ harmonics.
These observations may further support the hydrodynamic origin of collective correlations
in high-multiplicity events for these small systems~\cite{ALICE:2016kpq}.

However, the nature of the long-range collectivity in small systems, especially for the low-multiplicity region (e.g., less than about 50--60 
charged particles), still remains inconclusive and much debated (e.g., see reviews in Refs.~\cite{Dusling:2015gta,Nagle:2018nvi}). 
It has been argued that the contribution of initial momentum space 
collectivity from the gluon saturation model may become dominant as the event multiplicity decreases~\cite{Schenke:2019pmk}. 
Understanding the multiplicity dependence of the observed long-range collectivity is the key to disentangle contributions from various physical origins.
Experimental investigation of collective multiparticle correlations for low-multiplicity events is largely hindered by
the presence of significant noncollective correlations (nonflow), such as few-particle correlations from jets.
The observed trend for the $v_2$--$v_3$ correlation ($\text{SC}(n,m)$) to become positive is likely related to the nonflow effect~\cite{Sirunyan:2017uyl}.
In order to suppress these few-particle correlations and to explore possible collective correlation signals, 
subevent cumulant techniques have been proposed to require rapidity gaps among particles~\cite{DiFrancesco:2016srj,Jia:2017hbm}.
As detailed in Refs.~\cite{Jia:2017hbm,Aaboud:2017blb,Huo:2017nms}, each combination of four particles is required to fall into two,
three or four distinct subevents within the full $\eta$ range. There are already studies highlighting the importance of the 
nonflow contribution in cumulant calculations and the effectiveness of the subevent techniques to strongly suppress 
it~\cite{Aaboud:2017blb,Huo:2017nms}.

Using a large data sample collected using the CMS detector, this paper presents the first measurement of 
event-by-event correlations of $v_2$ \vs $v_3$ and $v_2$ \vs $v_4$ using the SC method with subevents 
in \pPb collisions at a nucleon-nucleon center-of mass energy $\sqrtsNN = 8.16\TeV$ covering a wide multiplicity range. 
The correlation measurements are performed using 2, 3, and 4 subevents, where the impact of few-particle correlations 
is systematically reduced in a data-driven way as the number of subevents increases. The results are also compared to previous measurements 
without the subevent technique.

\section{The CMS detector}

The central feature of the CMS apparatus is a superconducting solenoid of 6\unit{m}
internal diameter, providing a magnetic field of 3.8\unit{T}. Within the solenoid volume,
there are four primary subdetectors including a silicon
pixel and strip tracker detector, a lead tungstate crystal electromagnetic calorimeter
(ECAL), and a brass and scintillator hadron calorimeter (HCAL), each composed of a
barrel and two endcap sections. The iron and quartz-fiber Cherenkov hadron forward
(HF) calorimeters cover the range $3 < \abs{\eta} < 5$.
The silicon tracker measures charged particles within the range $\abs{\eta}< 2.5$. For charged
particles with transverse momentum $1 < \pt < 10\GeVc$ and $\abs{\eta} < 1.4$, the track resolutions are
typically 1.5\% in \pt and 25--90 (45--150)\mum in the transverse (longitudinal)
impact parameter~\cite{Chatrchyan:2014fea}. The Monte Carlo (MC) simulation of the full CMS detector response is based on
\GEANTfour~\cite{GEANT4}. The detailed description of the CMS detector can be found in Ref.~\cite{Chatrchyan:2008zzk}.

\section{Event and track selections}

The measurements presented in this paper use the 8.16\TeV \pPb data set with an integrated luminosity 
of 186\nbinv, where the beam directions were reversed during the run after collecting the first 62.6\nbinv. 
The beam energies were 6.5\TeV for protons and 2.56\TeV per nucleon for lead nuclei~\cite{PhysRevAccelBeams.20.081003}.
The results from both beam directions are combined using the convention that the proton-going direction 
defines positive pseudorapidity. As a result of the energy difference between the colliding beams, the nucleon-nucleon 
center-of-mass frame in the \pPb collisions is not at rest with respect to the laboratory frame.
Massless particles emitted at $\eta_{\text{CM}} = 0$ in the nucleon-nucleon center-of-mass frame will be 
detected at $\eta_{\text{lab}} = 0.465$ in the laboratory frame. All pseudorapidities reported in this paper are 
given with respect to the laboratory frame. During the data taking, the average number of collisions per bunch
crossing (pileup) varied from 0.10 to 0.25. A procedure similar to that described in Ref.~\cite{Chatrchyan:2013nka} 
is used for identifying and rejecting events with pileup.

The minimum bias (MB) 8.16\TeV \pPb events are triggered by requiring energy
deposits in at least one of the two HF calorimeters above 1\GeV and
the presence of at least one track with $\pt > 0.4$\GeVc reconstructed using hits from the pixel tracker only.
In order to collect a large sample of high-multiplicity \pPb collisions, a dedicated
trigger is implemented using the CMS level-1 (L1) and high-level
trigger (HLT) systems~\cite{Khachatryan:2016bia}. At L1, the total number of ECAL+HCAL towers having 
deposited energy above an energy threshold of 0.5\GeV in transverse energy (\ET)
is required to be greater than a given threshold (120 and 150 towers depending on the targeted multiplicity range).
As part of the HLT trigger, the track reconstruction is performed online with
the identical reconstruction algorithm used offline~\cite{Chatrchyan:2014fea}.
For each event selected at L1, the reconstructed vertex with the highest number of associated tracks is selected as the primary vertex at the HLT.
The number of tracks with $\abs{\eta}<2.4$, $\pt > 0.4\GeVc$, and a distance of closest
approach less than 0.12\unit{cm} along the beam axis to the primary vertex is determined for each event and is required
to exceed 120, 185 and 250 to enrich the sample with high-multiplicity (HM) events in the ranges 120--185, 185--250 and 250--$\infty$, respectively.
The events are required to contain a primary vertex within 15\unit{cm} of the nominal interaction point along the beam axis and 0.2\unit{cm} in the
transverse direction. Finally, for high-multiplicity events, the trigger efficiency is required to be greater than 95\%. In the multiplicity region where
this requirement is not met ($\noff < 120$), MB triggered events are used.

In the offline analysis, the primary tracks, i.e. reconstructed tracks that originate from the primary
vertex and satisfy the high-quality criteria of Ref.~\cite{Chatrchyan:2014fea},
are used to perform the correlation measurements, as well as to evaluate the charged-particle multiplicity (\noff) for each event.
In addition, the significances of the track impact parameter with respect to the primary vertex 
in the transverse and longitudinal direction divided by their uncertainties are required to be less than 3.
The relative \pt uncertainty must be less than 10\%. To ensure high tracking efficiency, only tracks with
$\abs{\eta}<2.4$ and $\pt > 0.3\GeVc$ are used in this analysis~\cite{Chatrchyan:2014fea}. 

In this analysis, about 8 billion MB and 500 million HM events are selected. Following the convention 
established in previous analyses~\cite{Khachatryan:2016got,Sirunyan:2017uyl,Sirunyan:2017quh}, the \pPb data are shown 
in classes of \noff, which is the number of primary tracks with $\abs{\eta}<2.4$ and $\pt >0.4$\GeVc, without corrections for acceptance and efficiency. 
The \noff boundaries used for the results of this paper are: 10, 20, 40, 80, 120, 150, 185, 250, and 350. 
These boundaries are chosen to minimize the statistical uncertainty in each bin. The average \noff 
for MB pPb events is about 40. The overall CMS acceptance and tracking efficiency is about 85\%. 

\section{Analysis technique}
\label{sec:analysis}

The SC technique, first introduced in Ref.~\cite{ALICE:2016kpq}, is based on four-particle correlations using cumulants.
The four-particle cumulant technique, by simultaneously correlating four particles, is known to have the advantage of suppressing nonflow quite
efficiently compared to other methods~\cite{Bilandzic:2013kga,Khachatryan:2016txc}. To study the correlation between the Fourier coefficients $n$ and $m$, one can build, for
each event, a 2-particle correlator ($\langle \cos (n\phi_1 - n\phi_2) \rangle$) and a 4-particle correlator ($\langle \cos (n\phi_1 + m\phi_2 - n\phi_3 - m\phi_4) \rangle$)
with a complex notation average over all the events as:
\ifthenelse{\boolean{cms@external}}{
\begin{linenomath}
\begin{equation}
\begin{aligned}
\dmean{2_{n,-n}} & \equiv \left<\left<\re^{i(n\phi_{1} - n\phi_{2})} \right>\right>, \\
\dmean{4_{n,m,-n,-m}} & \equiv \left<\left< \re^{i(n\phi_{1} + m\phi_{2}-n\phi_{3}-m\phi_{4})} \right>\right>.
\end{aligned}
\label{4pCorrelationAllEvent}
\end{equation}
\end{linenomath}
}{
\begin{linenomath}
\begin{equation}
\begin{aligned}
\dmean{2_{n,-n}} & \equiv \left<\left<\re^{i(n\phi_{1} - n\phi_{2})} \right>\right>, \\
\dmean{4_{n,m,-n,-m}} & \equiv \left<\left< \re^{i(n\phi_{1} + m\phi_{2}-n\phi_{3}-m\phi_{4})} \right>\right>.
\end{aligned}
\label{4pCorrelationAllEvent}
\end{equation}
\end{linenomath}
}
In the above equations, the real part of the 2- and 4-particle correlators are the cosine terms presented in Eq.~(\ref{scdef}.) The final observable, the $SC$, is defined as follows:
\begin{linenomath}
\begin{equation}
\label{eq:Fig79}
\text{SC}(n,m) = \dmean{4_{n,m,-n,-m}} - \dmean{2_{n,-n}} \, \dmean{2_{m,-m}}.
\end{equation}
\end{linenomath}
Nevertheless, it was shown in previous studies~\cite{Sirunyan:2017uyl} that the standard four-particle cumulant technique does not suppress all
of the short-range correlation contribution. In particular, the increasing trend of SC toward low multiplicities, following a power law, is characteristic of
remaining nonflow contaminations~\cite{Ollitrault:2009ie}.
In that paper, to further suppress nonflow, the subevent technique is used based on the calculation published in Ref.~\cite{DiFrancesco:2016srj}.
In the two-subevent case, the first and second subevents are defined as $-2.4 < \eta < 0$ and $0 < \eta < 2.4$. The bounds for three subevents are
$-$2.4, $-$0.8, 0.8, 2.4, and for four subevents are $-$2.4, $-$1.2, 0, 1.2, 2.4. The formula of the SC calculation can be derived from Eq.~(\ref{eq:Fig79}):
\ifthenelse{\boolean{cms@external}}{
\begin{align}
\text{SC}_{\text{2sub}}(n,m) &= \langle \langle 4 \rangle^{aa|bb}_{n,m|-n,-m} \rangle \nonumber \\
                             &- \langle \langle 2 \rangle^{a|b}_{n|-n} \rangle \langle \langle 2 \rangle^{a|b}_{m|-m} \rangle, \label{eq:subSC1}\\
\text{SC}_{\text{3sub}}(n,m) &= \langle \langle 4 \rangle^{a|bb|c}_{-n|m,n|-m} \rangle \nonumber \\
                             &- \langle \langle 2 \rangle^{a|b}_{-n|n} \rangle \langle \langle 2 \rangle^{b|c}_{m|-m} \rangle, \label{eq:subSC2}\\
\text{SC}_{\text{4sub}}(n,m) &= \langle \langle 4 \rangle^{a|b|c|d}_{n|m|-n|-m} \rangle \nonumber \\
                             &- \langle \langle 2 \rangle^{a|c}_{n|-n} \rangle \langle \langle 2 \rangle^{b|d}_{m|-m} \rangle. \label{eq:subSC3}
\end{align}
}{
\begin{align}
\text{SC}_{\text{2sub}}(n,m) &= \langle \langle 4 \rangle^{aa|bb}_{n,m|-n,-m} \rangle - \langle \langle 2 \rangle^{a|b}_{n|-n} \rangle \langle \langle 2 \rangle^{a|b}_{m|-m} \rangle, \label{eq:subSC1}\\
\text{SC}_{\text{3sub}}(n,m) &= \langle \langle 4 \rangle^{a|bb|c}_{-n|m,n|-m} \rangle - \langle \langle 2 \rangle^{a|b}_{-n|n} \rangle \langle \langle 2 \rangle^{b|c}_{m|-m} \rangle, \label{eq:subSC2}\\
\text{SC}_{\text{4sub}}(n,m) &= \langle \langle 4 \rangle^{a|b|c|d}_{n|m|-n|-m} \rangle - \langle \langle 2 \rangle^{a|c}_{n|-n} \rangle \langle \langle 2 \rangle^{b|d}_{m|-m} \rangle. \label{eq:subSC3}
\end{align}
}
\noindent where $a, b, c$, and $d$ denote the particles chosen in each subevent for the calculation and $n, m$ the corresponding harmonic attributed to this subevent.
In Eq.~(\ref{eq:subSC1}), the notation $aa|bb$ in the 4-particle correlator means that two particles are required to be in the first subevent ($aa$)
while the other two are required to be in the second subevent ($bb$). Similarly, for the 2-particle correlator, one particle in each subevent is required ($a|b$). A similar reasoning is
applied in Eqs.~(\ref{eq:subSC2}) and ~(\ref{eq:subSC3}).

The systematic uncertainties in the experimental procedure are evaluated by varying the conditions in
extracting SC. The systematic uncertainties due to tracking inefficiency and misreconstructed track rate are studied by varying the track quality requirements.
The selection thresholds on the significance of the transverse and longitudinal track impact parameter divided by their uncertainties are varied from 2 to 5.
In addition, the relative \pt uncertainty is varied from 5 to 10\%.
The sensitivity of the results to the primary vertex position along the beam axis ($z_\text{vtx}$) is quantified by comparing results with
different $z_\text{vtx}$ selection: $\abs{z_\text{vtx}} < 3$\unit{cm} and $3 < \abs{z_\text{vtx}} < 15$\unit{cm}, and the possible contamination by residual pileup
interactions is studied by varying the pileup rejection criteria from no pileup rejection at all to selecting events with only one
reconstructed vertex. Finally, to study potential trigger biases, a comparison to high-multiplicity \pPb data for a given multiplicity range that were collected
by a lower-threshold trigger with 100\% efficiency is performed. This uncertainty is found
to be negligible, while the other systematic uncertainty sources have contributions of 1\% each, independent of \noff.
The total systematic uncertainties are estimated to be 1.8\% for SC.

\section{Results}
\label{sec:results}

\begin{figure*}[hbt]
  \begin{center}
    \hspace{1cm}\includegraphics[width=.96\textwidth]{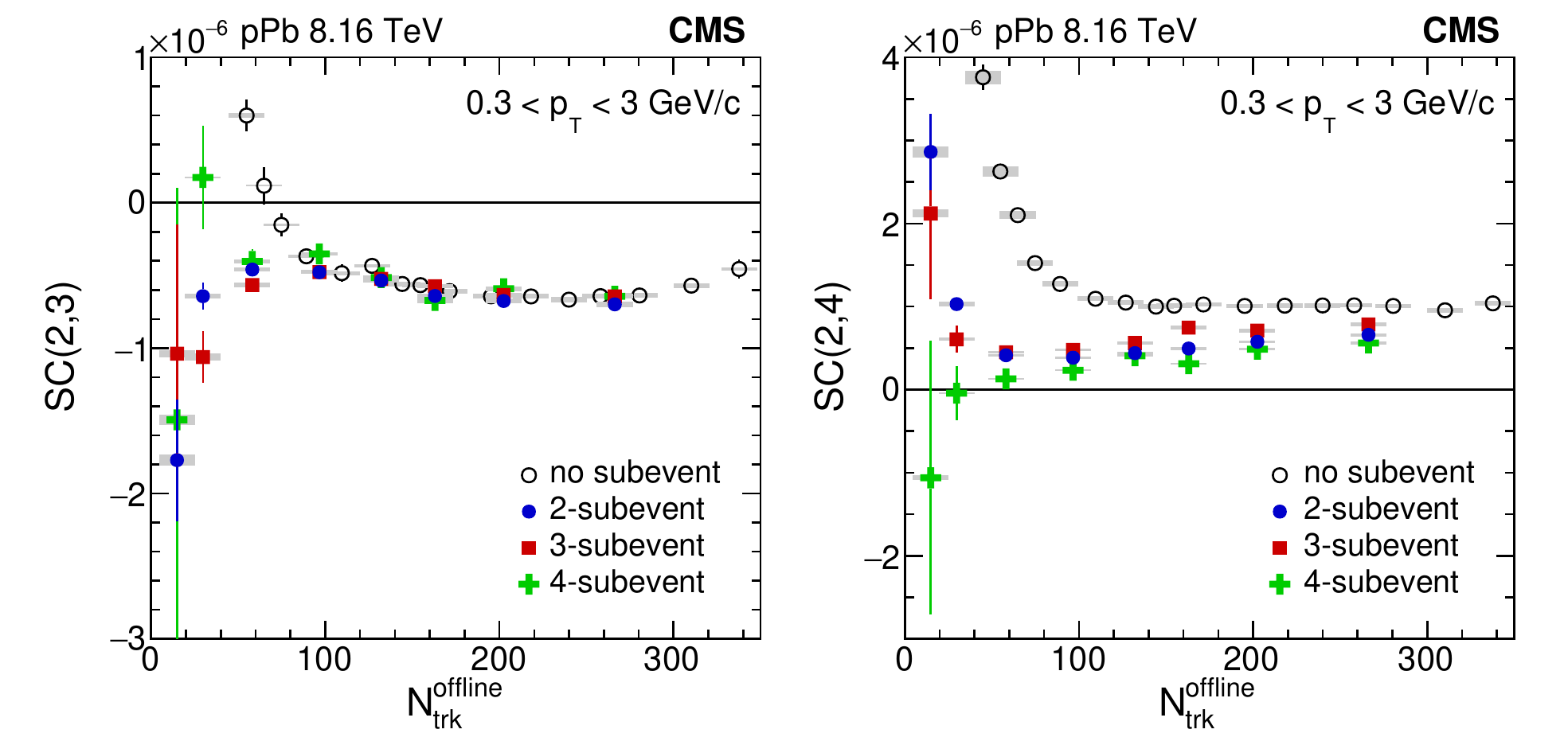}
    \caption{
The $\text{SC}(2,3)$ (left) and $\text{SC}(2,4)$ (right) distributions as functions of \noff from 2 subevents (full blue circles), 3 subevents (red squares), and 4 subevents (green crosses).
For comparison, published results from Ref.~\cite{Sirunyan:2017uyl} with no subevents (open black circles), are also shown.
Bars represent statistical uncertainties while grey areas represent the systematic uncertainties.}
    \label{fig:sc_all}
  \end{center}
\end{figure*}

The results of symmetric cumulants $\text{SC}(2,3)$ and $\text{SC}(2,4)$ obtained with the 2-, 3-, and 4-subevent 
methods for $0.3<\pt<3$\GeVc are shown in Fig.~\ref{fig:sc_all}, as functions of multiplicity in \pPb collisions at $\sqrtsNN = 8.16\TeV$.
For comparison, the results with no subevents from Ref.~\cite{Sirunyan:2017uyl} are also shown for the range $40 < \noff < 350$ 
(the SC with no subevents for lower multiplicities are out of range because of the choice of the y-axis scale). 
The systematic uncertainties are the same for no and $n$-subevents ($n = 2,3,4$).

Both $\text{SC}(2,3)$ and $\text{SC}(2,4)$ diverge toward large positive values for low-\noff 
ranges ($\noff < 80$) using the no-subevent method, likely because of a dominant contribution from few-particle short-range 
correlations, as discussed in Ref.~\cite{Sirunyan:2017uyl}. Using the subevent method, the contributions from short-range 
correlations are significantly suppressed~\cite{Aaboud:2017blb,Huo:2017nms}. 
No significant positive $\text{SC}(2,3)$ values with subevent methods are observed over the entire event multiplicity range. 
The 2- and 3-subevent $\text{SC}(2,3)$ preserve significant negative signals down to $\noff \sim 20$, while the 4-subevent 
$\text{SC}(2,3)$ tends to show a monotonic trend gradually converging to zero at $\noff \sim 20$. Similar
behavior is also observed for $\text{SC}(2,4)$, where 2- and 3-subevent $\text{SC}(2,4)$ values remain positive but the 
4-subevent $\text{SC}(2,4)$ decreases to zero toward $\noff \sim 20$. As the 4-subevent method is the most powerful in
eliminating nonflow effects, the observed trends in 4-subevent $\text{SC}(2,3)$ and  $\text{SC}(2,4)$ provide
evidence for the onset of long-range collective particle correlations from low to high multiplicities in \pPb collisions.

\begin{figure*}[hbt]
  \begin{center}
    \hspace{1cm}\includegraphics[width=.96\textwidth]{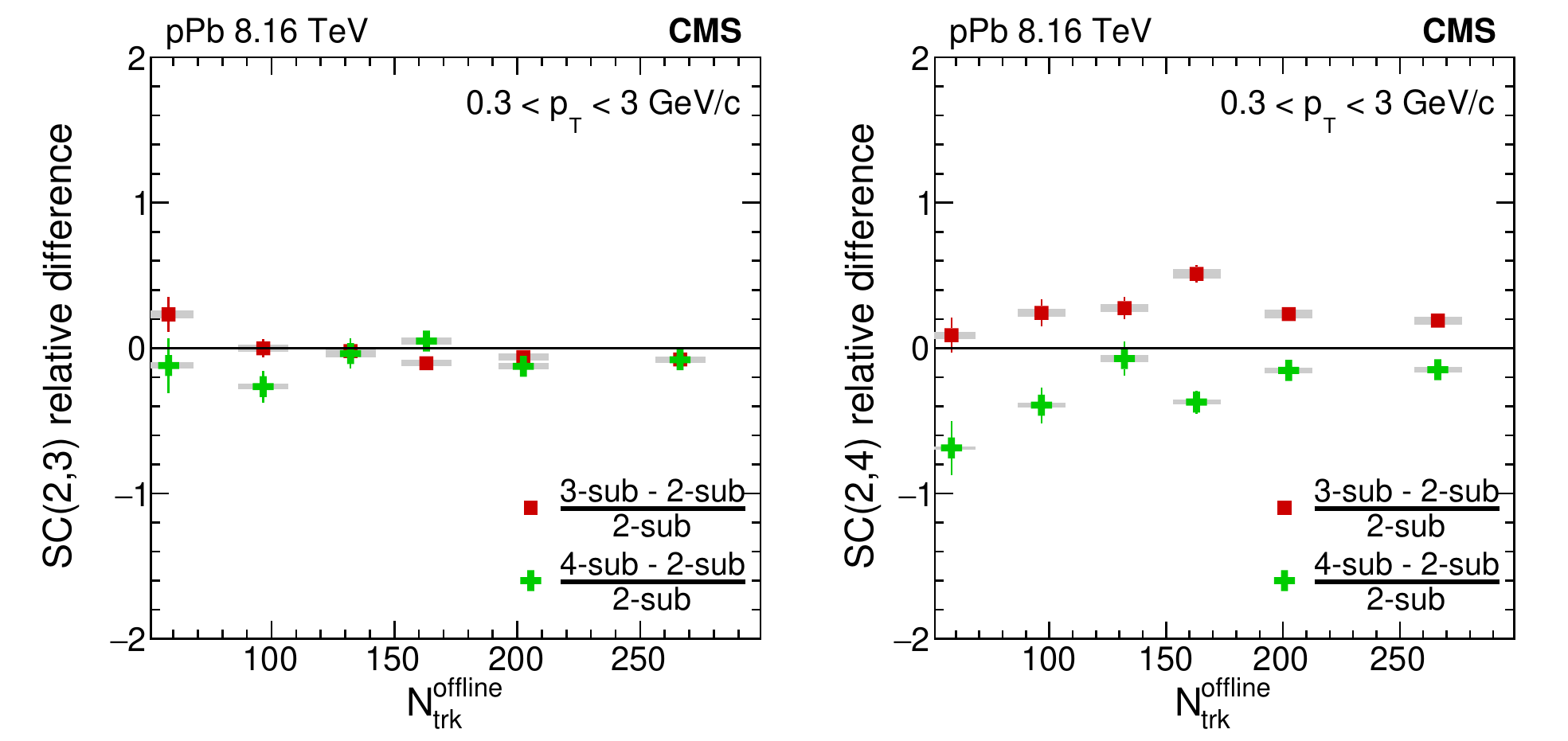}
    \caption{
The relative difference of $\text{SC}(2,3)$ (left) and $\text{SC}(2,4)$ (right) between 2 and 3 subevents (red squares) as well as between 2 and 4 subevents (green crosses) as a function of \noff.
Bars represent statistical uncertainties while shaded areas represent the systematic uncertainties.}
    \label{fig:sc_ratio}
  \end{center}
\end{figure*}

For $\noff > 80$, the no-subevent and $n$-subevent methods give consistent results for $\text{SC}(2,3)$, suggesting 
that the contribution from nonflow effects is negligible. For $\text{SC}(2,4)$, there is a difference clearly observed between no-subevent 
and $n$-subevent results even up to the highest multiplicities investigated.
This observation is illustrated more clearly in Fig.~\ref{fig:sc_ratio}, which shows the $\text{SC}(2,3)$ and $\text{SC}(2,4)$ 
relative differences between 2 subevents and 3 or 4 subevents. 
The $\text{SC}(2,3)$ results (Fig.~\ref{fig:sc_ratio}, left) are consistent among the 2-, 3- and 4-subevent methods, while there is an approximately
10--40\% difference for $\text{SC}(2,4)$ (Fig.~\ref{fig:sc_ratio}, right)  between the 2-subevent and 3- or 4-subevent methods. 
The 3-subevent $\text{SC}(2,4)$ values are greater than the 2-subevent values, contrary to what is typically 
expected from non-flow contributions. This behavior may suggest the sensitivity of $\text{SC}(2,4)$ to other effects.
In particular, the event-plane decorrelation~\cite{Khachatryan:2015oea} could be an important contribution to the 
observed behavior as also observed in Ref.~\cite{Aaboud:2018syf}. The impact of event-plane decorrelation and how it may be
different for $\text{SC}(2,3)$ and $\text{SC}(2,4)$ remains to be understood in future work. 

\section{Summary}
\label{sec:summary}

The first measurement of event-by-event correlations of different Fourier harmonic orders in symmetric 
cumulants $\text{SC}(2,3)$ and $\text{SC}(2,4)$ with 2, 3, and 4 subevents in proton-lead (\pPb) collisions at $\sqrtsNN = 8.16\TeV$ is
presented using a large data sample collected by the CMS experiment. The \pPb data analyzed with the subevent method 
are compared to previously published results using the technique without subevents. In all cases, an anticorrelation is observed 
between the single-particle anisotropy harmonics $v_2$ and $v_3$, while $v_2$ and $v_4$ are positively correlated.
For charged-particle multiplicity $\noff > 100$, both standard and $n$-subevent methods give similar results for 
$\text{SC}(2,3)$, suggesting that nonflow effects have negligible contributions in this region. The $\text{SC}(2,4)$
results show a somewhat different behavior, which depends on the number of subevents in the same multiplicity region. 
By significantly suppressing the nonflow contribution, the 4-subevent results for both $\text{SC}(2,3)$ and $\text{SC}(2,4)$
show a monotonically decreasing magnitude toward zero at $\noff \sim 20$. These new results presented in this paper 
provide evidence for the onset of long-range collective particle correlations from low to high multiplicity events in \pPb collisions.
The observed multiplicity dependence of multiparticle azimuthal correlations may further constrain the physical origin 
of the collectivity observed in small system collisions.

\begin{acknowledgments}
We congratulate our colleagues in the CERN accelerator departments for the excellent performance of the LHC and thank the technical and administrative staffs at CERN and at other CMS institutes for their contributions
to the success of the CMS effort. In addition, we gratefully acknowledge the computing centres and personnel of the Worldwide LHC Computing Grid for delivering so effectively the computing infrastructure essential
to our analyses. Finally, we acknowledge the enduring support for the construction and operation of the LHC and the CMS detector provided by the following funding agencies: BMBWF and FWF (Austria); FNRS and FWO
(Belgium); CNPq, CAPES, FAPERJ, FAPERGS, and FAPESP (Brazil); MES (Bulgaria); CERN; CAS, MoST, and NSFC (China); COLCIENCIAS (Colombia); MSES and CSF (Croatia); RPF (Cyprus); SENESCYT (Ecuador); MoER, ERC IUT, PUT
and ERDF (Estonia); Academy of Finland, MEC, and HIP (Finland); CEA and CNRS/IN2P3 (France); BMBF, DFG, and HGF (Germany); GSRT (Greece); NKFIA (Hungary); DAE and DST (India); IPM (Iran); SFI (Ireland); INFN (Italy);
MSIP and NRF (Republic of Korea); MES (Latvia); LAS (Lithuania); MOE and UM (Malaysia); BUAP, CINVESTAV, CONACYT, LNS, SEP, and UASLP-FAI (Mexico); MOS (Montenegro); MBIE (New Zealand); PAEC (Pakistan); MSHE and NSC
(Poland); FCT (Portugal); JINR (Dubna); MON, RosAtom, RAS, RFBR, and NRC KI (Russia); MESTD (Serbia); SEIDI, CPAN, PCTI, and FEDER (Spain); MOSTR (Sri Lanka); Swiss Funding Agencies (Switzerland); MST (Taipei);
ThEPCenter, IPST, STAR, and NSTDA (Thailand); TUBITAK and TAEK (Turkey); NASU and SFFR (Ukraine); STFC (United Kingdom); DOE and NSF (USA).

\hyphenation{Rachada-pisek} Individuals have received support from the Marie-Curie programme and the European Research Council and Horizon 2020 Grant, contract Nos. 675440 and 765710 (European Union); the Leventis
Foundation; the A.P. Sloan Foundation; the Alexander von Humboldt Foundation; the Belgian Federal Science Policy Office; the Fonds pour la Formation \`a la Recherche dans l'Industrie et dans l'Agriculture
(FRIA-Belgium); the Agentschap voor Innovatie door Wetenschap en Technologie (IWT-Belgium); the F.R.S.-FNRS and FWO (Belgium) under the ``Excellence of Science -- EOS" -- be.h project n. 30820817; the Beijing
Municipal Science \& Technology Commission, No. Z181100004218003; the Ministry of Education, Youth and Sports (MEYS) of the Czech Republic; the Lend\"ulet (``Momentum") Programme and the J\'anos Bolyai Research
Scholarship of the Hungarian Academy of Sciences, the New National Excellence Program \'UNKP, the NKFIA research grants 123842, 123959, 124845, 124850, 125105, 128713, 128786, and 129058 (Hungary); the Council of
Science and Industrial Research, India; the HOMING PLUS programme of the Foundation for Polish Science, cofinanced from European Union, Regional Development Fund, the Mobility Plus programme of the Ministry of
Science and Higher Education, the National Science Center (Poland), contracts Harmonia 2014/14/M/ST2/00428, Opus 2014/13/B/ST2/02543, 2014/15/B/ST2/03998, and 2015/19/B/ST2/02861, Sonata-bis 2012/07/E/ST2/01406;
the National Priorities Research Program by Qatar National Research Fund; the Programa Estatal de Fomento de la Investigaci{\'o}n Cient{\'i}fica y T{\'e}cnica de Excelencia Mar\'{\i}a de Maeztu, grant MDM-2015-0509
and the Programa Severo Ochoa del Principado de Asturias; the Thalis and Aristeia programmes cofinanced by EU-ESF and the Greek NSRF; the Rachadapisek Sompot Fund for Postdoctoral Fellowship, Chulalongkorn University
and the Chulalongkorn Academic into Its 2nd Century Project Advancement Project (Thailand); the Welch Foundation, contract C-1845; and the Weston Havens Foundation (USA).
\end{acknowledgments}
\bibliography{auto_generated}

\cleardoublepage \appendix\section{The CMS Collaboration \label{app:collab}}\begin{sloppypar}\hyphenpenalty=5000\widowpenalty=500\clubpenalty=5000\input{HIN-18-015-authorlist.tex}\end{sloppypar}
\end{document}

%% file: HIN-18-015-authorlist.tex
\vskip\cmsinstskip
\textbf{Yerevan Physics Institute, Yerevan, Armenia}\\*[0pt]
A.M.~Sirunyan, A.~Tumasyan
\vskip\cmsinstskip
\textbf{Institut f\"{u}r Hochenergiephysik, Wien, Austria}\\*[0pt]
W.~Adam, F.~Ambrogi, E.~Asilar, T.~Bergauer, J.~Brandstetter, M.~Dragicevic, J.~Er\"{o}, A.~Escalante~Del~Valle, M.~Flechl, R.~Fr\"{u}hwirth\cmsAuthorMark{1}, V.M.~Ghete, J.~Hrubec, M.~Jeitler\cmsAuthorMark{1}, N.~Krammer, I.~Kr\"{a}tschmer, D.~Liko, T.~Madlener, I.~Mikulec, N.~Rad, H.~Rohringer, J.~Schieck\cmsAuthorMark{1}, R.~Sch\"{o}fbeck, M.~Spanring, D.~Spitzbart, A.~Taurok, W.~Waltenberger, J.~Wittmann, C.-E.~Wulz\cmsAuthorMark{1}, M.~Zarucki
% ...
\vskip\cmsinstskip
% ...
1:  Also at Vienna University of Technology, Vienna, Austria\\
% ...


%% file: HIN-18-015_temp.bbl
\providecommand{\href}[2]{#2}\begingroup\raggedright\begin{thebibliography}{10}%
\makeatletter
\providecommand{\hrefCMSnoop }[0]{\@secondoftwo}%
\makeatother
\providecommand{\doi}{\texttt{doi:}\begingroup \urlstyle{tt}\Url}

\bibitem{Shuryak:2004cy}
\hrefCMSnoop {}{E.~V. Shuryak, ``{What RHIC experiments and theory tell us
  about properties of quark-gluon plasma?}'',} \textit{ Nucl. Phys. A} \textbf{
  750} (2005) 64,
  \href{http://dx.doi.org/10.1016/j.nuclphysa.2004.10.022}{\doi{10.1016/j.nuclphysa.2004.10.022}},
\href{http://www.arXiv.org/abs/hep-ph/0405066}{\texttt{arXiv:hep-ph/0405066}}.

\bibitem{Busza:2018rrf}
\hrefCMSnoop {}{W.~Busza, K.~Rajagopal, and W.~van~der Schee, ``{Heavy ion
  collisions: The big picture, and the big questions}'',} \textit{ Ann. Rev.
  Nucl. Part. Sci.} \textbf{ 68} (2018) 339,
  \href{http://dx.doi.org/10.1146/annurev-nucl-101917-020852}{\doi{10.1146/annurev-nucl-101917-020852}},
\href{http://www.arXiv.org/abs/1802.04801}{\texttt{arXiv:1802.04801}}.

\bibitem{Abelev:2009af}
\hrefCMSnoop {}{{STAR} Collaboration, ``{Long range rapidity correlations and
  jet production in high energy nuclear collisions}'',} \textit{ Phys. Rev. C}
  \textbf{ 80} (2009) 064912,
  \href{http://dx.doi.org/10.1103/PhysRevC.80.064912}{\doi{10.1103/PhysRevC.80.064912}},
\href{http://www.arXiv.org/abs/0909.0191}{\texttt{arXiv:0909.0191}}.

\bibitem{Alver:2008gk}
\hrefCMSnoop {}{{PHOBOS} Collaboration, ``{System size dependence of cluster
  properties from two- particle angular correlations in Cu+Cu and Au+Au
  collisions at \sqrtsNN\ = 200\GeV}'',} \textit{ Phys. Rev. C} \textbf{ 81}
  (2010) 024904,
  \href{http://dx.doi.org/10.1103/PhysRevC.81.024904}{\doi{10.1103/PhysRevC.81.024904}},
\href{http://www.arXiv.org/abs/0812.1172}{\texttt{arXiv:0812.1172}}.

\bibitem{Alver:2009id}
\hrefCMSnoop {}{{PHOBOS} Collaboration, ``{High transverse momentum triggered
  correlations over a large pseudorapidity acceptance in Au+Au collisions at
  \sqrtsNN\ = 200\GeV}'',} \textit{ Phys. Rev. Lett.} \textbf{ 104} (2010)
  062301,
  \href{http://dx.doi.org/10.1103/PhysRevLett.104.062301}{\doi{10.1103/PhysRevLett.104.062301}},
\href{http://www.arXiv.org/abs/0903.2811}{\texttt{arXiv:0903.2811}}.

\bibitem{Abelev:2009jv}
\hrefCMSnoop {}{{STAR} Collaboration, ``{Three-particle coincidence of the long
  range pseudorapidity correlation in high energy nucleus-nucleus
  collisions}'',} \textit{ Phys. Rev. Lett.} \textbf{ 105} (2010) 022301,
  \href{http://dx.doi.org/10.1103/PhysRevLett.105.022301}{\doi{10.1103/PhysRevLett.105.022301}},
\href{http://www.arXiv.org/abs/0912.3977}{\texttt{arXiv:0912.3977}}.

\bibitem{Chatrchyan:2011eka}
\hrefCMSnoop {}{{CMS Collaboration}, ``{Long-range and short-range dihadron
  angular correlations in central PbPb collisions at a nucleon-nucleon center
  of mass energy of 2.76 TeV}'',} \textit{ JHEP} \textbf{ 07} (2011) 076,
  \href{http://dx.doi.org/10.1007/JHEP07(2011)076}{\doi{10.1007/JHEP07(2011)076}},
\href{http://www.arXiv.org/abs/1105.2438}{\texttt{arXiv:1105.2438}}.

\bibitem{Chatrchyan:2012wg}
\hrefCMSnoop {}{{CMS Collaboration}, ``{Centrality dependence of dihadron
  correlations and azimuthal anisotropy harmonics in PbPb collisions at
  \sqrtsNN\ = 2.76 TeV}'',} \textit{ Eur. Phys. J. C} \textbf{ 72} (2012) 2012,
  \href{http://dx.doi.org/10.1140/epjc/s10052-012-2012-3}{\doi{10.1140/epjc/s10052-012-2012-3}},
\href{http://www.arXiv.org/abs/1201.3158}{\texttt{arXiv:1201.3158}}.

\bibitem{Aamodt:2011by}
\hrefCMSnoop {}{{ALICE Collaboration}, ``{Harmonic decomposition of
  two-particle angular correlations in Pb-Pb collisions at \sqrtsNN =
  2.76\TeV}'',} \textit{ Phys. Lett. B} \textbf{ 708} (2012) 249,
  \href{http://dx.doi.org/10.1016/j.physletb.2012.01.060}{\doi{10.1016/j.physletb.2012.01.060}},
\href{http://www.arXiv.org/abs/1109.2501}{\texttt{arXiv:1109.2501}}.

\bibitem{ATLAS:2012at}
\hrefCMSnoop {}{{ATLAS Collaboration}, ``{Measurement of the azimuthal
  anisotropy for charged particle production in \sqrtsNN\ = 2.76\TeV lead-lead
  collisions with the ATLAS detector}'',} \textit{ Phys. Rev. C} \textbf{ 86}
  (2012) 014907,
  \href{http://dx.doi.org/10.1103/PhysRevC.86.014907}{\doi{10.1103/PhysRevC.86.014907}},
\href{http://www.arXiv.org/abs/1203.3087}{\texttt{arXiv:1203.3087}}.

\bibitem{CMS:2013bza}
\hrefCMSnoop {}{{CMS Collaboration}, ``{Studies of azimuthal dihadron
  correlations in ultra-central PbPb collisions at \sqrtsNN\ = 2.76\TeV}'',}
  \textit{ JHEP} \textbf{ 02} (2014) 088,
  \href{http://dx.doi.org/10.1007/JHEP02(2014)088}{\doi{10.1007/JHEP02(2014)088}},
\href{http://www.arXiv.org/abs/1312.1845}{\texttt{arXiv:1312.1845}}.

\bibitem{Ollitrault:1992bk}
\hrefCMSnoop {}{J.-Y. Ollitrault, ``{Anisotropy as a signature of transverse
  collective flow}'',} \textit{ Phys. Rev. D} \textbf{ 46} (1992) 229,
\href{http://dx.doi.org/10.1103/PhysRevD.46.229}{\doi{10.1103/PhysRevD.46.229}}.

\bibitem{Alver:2010gr}
\hrefCMSnoop {}{B.~Alver and G.~Roland, ``{Collision geometry fluctuations and
  triangular flow in heavy-ion collisions}'',} \textit{ Phys. Rev. C} \textbf{
  81} (2010) 054905,
  \href{http://dx.doi.org/10.1103/PhysRevC.81.054905}{\doi{10.1103/PhysRevC.81.054905}},
  \href{http://www.arXiv.org/abs/1003.0194}{\texttt{arXiv:1003.0194}}.
[Erratum: \DOI{10.1103/PhysRevC.82.039903}].

\bibitem{Voloshin:1994mz}
\hrefCMSnoop {}{S.~Voloshin and Y.~Zhang, ``{Flow study in relativistic nuclear
  collisions by Fourier expansion of azimuthal particle distributions}'',}
  \textit{ Z. Phys. C} \textbf{ 70} (1996) 665,
  \href{http://dx.doi.org/10.1007/s002880050141}{\doi{10.1007/s002880050141}},
\href{http://www.arXiv.org/abs/hep-ph/9407282}{\texttt{arXiv:hep-ph/9407282}}.

\bibitem{Aad:2015lwa}
\hrefCMSnoop {}{{ATLAS Collaboration}, ``{Measurement of the correlation
  between flow harmonics of different order in lead-lead collisions at
  \sqrtsNN\ = 2.76 TeV with the ATLAS detector}'',} \textit{ Phys. Rev. C}
  \textbf{ 92} (2015) 034903,
  \href{http://dx.doi.org/10.1103/PhysRevC.92.034903}{\doi{10.1103/PhysRevC.92.034903}},
\href{http://www.arXiv.org/abs/1504.01289}{\texttt{arXiv:1504.01289}}.

\bibitem{ALICE:2016kpq}
\hrefCMSnoop {}{{ALICE Collaboration}, ``{Correlated event-by-event
  fluctuations of flow harmonics in Pb-Pb collisions at \sqrtsNN\ = 2.76
  TeV}'',} \textit{ Phys. Rev. Lett.} \textbf{ 117} (2016) 182301,
  \href{http://dx.doi.org/10.1103/PhysRevLett.117.182301}{\doi{10.1103/PhysRevLett.117.182301}},
\href{http://www.arXiv.org/abs/1604.07663}{\texttt{arXiv:1604.07663}}.

\bibitem{Bilandzic:2013kga}
A.~Bilandzic\hrefCMSnoop {}{ {et~al.}, ``{Generic framework for anisotropic
  flow analyses with multiparticle azimuthal correlations}'',} \textit{ Phys.
  Rev. C} \textbf{ 89} (2014) 064904,
  \href{http://dx.doi.org/10.1103/PhysRevC.89.064904}{\doi{10.1103/PhysRevC.89.064904}},
\href{http://www.arXiv.org/abs/1312.3572}{\texttt{arXiv:1312.3572}}.

\bibitem{Alver:2010dn}
\hrefCMSnoop {}{B.~H. Alver, C.~Gombeaud, M.~Luzum, and J.-Y. Ollitrault,
  ``{Triangular flow in hydrodynamics and transport theory}'',} \textit{ Phys.
  Rev. C} \textbf{ 82} (2010) 034913,
  \href{http://dx.doi.org/10.1103/PhysRevC.82.034913}{\doi{10.1103/PhysRevC.82.034913}},
\href{http://www.arXiv.org/abs/1007.5469}{\texttt{arXiv:1007.5469}}.

\bibitem{Schenke:2010rr}
\hrefCMSnoop {}{B.~Schenke, S.~Jeon, and C.~Gale, ``Elliptic and triangular
  flow in event-by-event {D=3+1} viscous hydrodynamics'',} \textit{ Phys. Rev.
  Lett.} \textbf{ 106} (2011) 042301,
  \href{http://dx.doi.org/10.1103/PhysRevLett.106.042301}{\doi{10.1103/PhysRevLett.106.042301}},
\href{http://www.arXiv.org/abs/1009.3244}{\texttt{arXiv:1009.3244}}.

\bibitem{Qiu:2011hf}
\hrefCMSnoop {}{Z.~Qiu, C.~Shen, and U.~Heinz, ``{Hydrodynamic elliptic and
  triangular flow in Pb-Pb collisions at \sqrtsNN\ = 2.76\TeV}'',} \textit{
  Phys. Lett. B} \textbf{ 707} (2012) 151,
  \href{http://dx.doi.org/10.1016/j.physletb.2011.12.041}{\doi{10.1016/j.physletb.2011.12.041}},
\href{http://www.arXiv.org/abs/1110.3033}{\texttt{arXiv:1110.3033}}.

\bibitem{Giacalone:2016afq}
\hrefCMSnoop {}{G.~Giacalone, L.~Yan, J.~Noronha-Hostler, and J.-Y. Ollitrault,
  ``{Symmetric cumulants and event-plane correlations in Pb + Pb
  collisions}'',} \textit{ Phys. Rev. C} \textbf{ 94} (2016) 014906,
  \href{http://dx.doi.org/10.1103/PhysRevC.94.014906}{\doi{10.1103/PhysRevC.94.014906}},
\href{http://www.arXiv.org/abs/1605.08303}{\texttt{arXiv:1605.08303}}.

\bibitem{Khachatryan:2010gv}
\hrefCMSnoop {}{{CMS Collaboration}, ``{Observation of long-range near-side
  angular correlations in proton-proton collisions at the LHC}'',} \textit{
  JHEP} \textbf{ 09} (2010) 091,
  \href{http://dx.doi.org/10.1007/JHEP09(2010)091}{\doi{10.1007/JHEP09(2010)091}},
\href{http://www.arXiv.org/abs/1009.4122}{\texttt{arXiv:1009.4122}}.

\bibitem{Khachatryan:2015lva}
\hrefCMSnoop {}{{CMS Collaboration}, ``{Measurement of long-range near-side
  two-particle angular correlations in pp collisions at $\sqrt s =$ 13 TeV}'',}
  \textit{ Phys. Rev. Lett.} \textbf{ 116} (2016) 172302,
  \href{http://dx.doi.org/10.1103/PhysRevLett.116.172302}{\doi{10.1103/PhysRevLett.116.172302}},
\href{http://www.arXiv.org/abs/1510.03068}{\texttt{arXiv:1510.03068}}.

\bibitem{Aad:2015gqa}
\hrefCMSnoop {}{{ATLAS Collaboration}, ``{Observation of long-range elliptic
  azimuthal anisotropies in $\sqrt{s}=$ 13 and 2.76 TeV pp collisions with the
  ATLAS detector}'',} \textit{ Phys. Rev. Lett.} \textbf{ 116} (2016) 172301,
  \href{http://dx.doi.org/10.1103/PhysRevLett.116.172301}{\doi{10.1103/PhysRevLett.116.172301}},
\href{http://www.arXiv.org/abs/1509.04776}{\texttt{arXiv:1509.04776}}.

\bibitem{Aad:2014lta}
\hrefCMSnoop {}{{ATLAS Collaboration}, ``{Measurement of long-range
  pseudorapidity correlations and azimuthal harmonics in \sqrtsNN\ = 5.02 TeV
  proton-lead collisions with the ATLAS detector}'',} \textit{ Phys. Rev. C}
  \textbf{ 90} (2014) 044906,
  \href{http://dx.doi.org/10.1103/PhysRevC.90.044906}{\doi{10.1103/PhysRevC.90.044906}},
\href{http://www.arXiv.org/abs/1409.1792}{\texttt{arXiv:1409.1792}}.

\bibitem{Khachatryan:2014jra}
\hrefCMSnoop {}{{CMS Collaboration}, ``{Long-range two-particle correlations of
  strange hadrons with charged particles in pPb and PbPb collisions at LHC
  energies}'',} \textit{ Phys. Lett. B} \textbf{ 742} (2015) 200,
  \href{http://dx.doi.org/10.1016/j.physletb.2015.01.034}{\doi{10.1016/j.physletb.2015.01.034}},
\href{http://www.arXiv.org/abs/1409.3392}{\texttt{arXiv:1409.3392}}.

\bibitem{Aaij:2015qcq}
\hrefCMSnoop {}{{LHCb Collaboration}, ``{Measurements of long-range near-side
  angular correlations in \sqrtsNN\ = 5 TeV proton-lead collisions in the
  forward region}'',} \textit{ Phys. Lett. B} \textbf{ 762} (2016) 473,
  \href{http://dx.doi.org/10.1016/j.physletb.2016.09.064}{\doi{10.1016/j.physletb.2016.09.064}},
\href{http://www.arXiv.org/abs/1512.00439}{\texttt{arXiv:1512.00439}}.

\bibitem{Adamczyk:2015xjc}
\hrefCMSnoop {}{{STAR} Collaboration, ``{Long-range pseudorapidity dihadron
  correlations in $d$+Au collisions at $\sqrt{s_{\rm NN}}=200$ GeV}'',}
  \textit{ Phys. Lett. B} \textbf{ 747} (2015) 265,
  \href{http://dx.doi.org/10.1016/j.physletb.2015.05.075}{\doi{10.1016/j.physletb.2015.05.075}},
\href{http://www.arXiv.org/abs/1502.07652}{\texttt{arXiv:1502.07652}}.

\bibitem{Khachatryan:2015waa}
\hrefCMSnoop {}{{CMS Collaboration}, ``{Evidence for collective multi-particle
  correlations in pPb collisions}'',} \textit{ Phys. Rev. Lett.} \textbf{ 115}
  (2015) 012301,
  \href{http://dx.doi.org/10.1103/PhysRevLett.115.012301}{\doi{10.1103/PhysRevLett.115.012301}},
\href{http://www.arXiv.org/abs/1502.05382}{\texttt{arXiv:1502.05382}}.

\bibitem{Khachatryan:2016txc}
\hrefCMSnoop {}{{CMS Collaboration}, ``{Evidence for collectivity in pp
  collisions at the LHC}'',} \textit{ Phys. Lett. B} \textbf{ 765} (2017) 193,
  \href{http://dx.doi.org/10.1016/j.physletb.2016.12.009}{\doi{10.1016/j.physletb.2016.12.009}},
\href{http://www.arXiv.org/abs/1606.06198}{\texttt{arXiv:1606.06198}}.

\bibitem{Aaboud:2017acw}
\hrefCMSnoop {}{{ATLAS Collaboration}, ``{Measurement of multi-particle
  azimuthal correlations in $pp$, $p+$Pb and low-multiplicity Pb$+$Pb
  collisions with the ATLAS detector}'',} \textit{ Eur. Phys. J. C} \textbf{
  77} (2017) 428,
  \href{http://dx.doi.org/10.1140/epjc/s10052-017-4988-1}{\doi{10.1140/epjc/s10052-017-4988-1}},
\href{http://www.arXiv.org/abs/1705.04176}{\texttt{arXiv:1705.04176}}.

\bibitem{Aaboud:2018syf}
\hrefCMSnoop {}{{ATLAS Collaboration}, ``{Correlated long-range mixed-harmonic
  fluctuations measured in $pp$, $p$+Pb and low-multiplicity Pb+Pb collisions
  with the ATLAS detector}'',} \textit{ Phys. Lett. B} \textbf{ 789} (2019)
  444,
  \href{http://dx.doi.org/10.1016/j.physletb.2018.11.065}{\doi{10.1016/j.physletb.2018.11.065}},
\href{http://www.arXiv.org/abs/1807.02012}{\texttt{arXiv:1807.02012}}.

\bibitem{Sirunyan:2017uyl}
\hrefCMSnoop {}{{CMS Collaboration}, ``Observation of correlated azimuthal
  anisotropy {Fourier} harmonics in pp and {p+Pb} collisions at the {LHC}'',}
  \textit{ Phys. Rev. Lett.} \textbf{ 120} (2018) 092301,
  \href{http://dx.doi.org/10.1103/PhysRevLett.120.092301}{\doi{10.1103/PhysRevLett.120.092301}},
\href{http://www.arXiv.org/abs/1709.09189}{\texttt{arXiv:1709.09189}}.

\bibitem{Dusling:2015gta}
\hrefCMSnoop {}{K.~Dusling, W.~Li, and B.~Schenke, ``{Novel collective
  phenomena in high-energy proton-proton and proton-nucleus collisions}'',}
  \textit{ Int. J. Mod. Phys. E} \textbf{ 25} (2016) 1630002,
  \href{http://dx.doi.org/10.1142/S0218301316300022}{\doi{10.1142/S0218301316300022}},
\href{http://www.arXiv.org/abs/1509.07939}{\texttt{arXiv:1509.07939}}.

\bibitem{Nagle:2018nvi}
\hrefCMSnoop {}{J.~L. Nagle and W.~A. Zajc, ``Small system collectivity in
  relativistic hadronic and nuclear collisions'',} \textit{ Ann. Rev. Nucl.
  Part. Sci.} \textbf{ 68} (2018) 211,
  \href{http://dx.doi.org/10.1146/annurev-nucl-101916-123209}{\doi{10.1146/annurev-nucl-101916-123209}},
  \href{http://www.arXiv.org/abs/1801.03477}{\texttt{arXiv:1801.03477}}.

\bibitem{Schenke:2019pmk}
\hrefCMSnoop {}{B.~Schenke, C.~Shen, and P.~Tribedy, ``Hybrid color glass
  condensate and hydrodynamic description of the {Relativistic Heavy Ion
  Collider} small system scan'',} \textit{ Phys. Lett. B} \textbf{ 803} (2020)
  135322,
  \href{http://dx.doi.org/10.1016/j.physletb.2020.135322}{\doi{10.1016/j.physletb.2020.135322}},
  \href{http://www.arXiv.org/abs/1908.06212}{\texttt{arXiv:1908.06212}}.

\bibitem{DiFrancesco:2016srj}
\hrefCMSnoop {}{P.~Di~Francesco, M.~Guilbaud, M.~Luzum, and J.-Y. Ollitrault,
  ``{Systematic procedure for analyzing cumulants at any order}'',} \textit{
  Phys. Rev. C} \textbf{ 95} (2017) 044911,
  \href{http://dx.doi.org/10.1103/PhysRevC.95.044911}{\doi{10.1103/PhysRevC.95.044911}},
\href{http://www.arXiv.org/abs/1612.05634}{\texttt{arXiv:1612.05634}}.

\bibitem{Jia:2017hbm}
\hrefCMSnoop {}{J.~Jia, M.~Zhou, and A.~Trzupek, ``{Revealing long-range
  multiparticle collectivity in small collision systems via subevent
  cumulants}'',} \textit{ Phys. Rev. C} \textbf{ 96} (2017) 034906,
  \href{http://dx.doi.org/10.1103/PhysRevC.96.034906}{\doi{10.1103/PhysRevC.96.034906}},
\href{http://www.arXiv.org/abs/1701.03830}{\texttt{arXiv:1701.03830}}.

\bibitem{Aaboud:2017blb}
\hrefCMSnoop {}{{ATLAS Collaboration}, ``{Measurement of long-range
  multiparticle azimuthal correlations with the subevent cumulant method in
  $pp$ and p+Pb collisions with the ATLAS detector at the CERN Large Hadron
  Collider}'',} \textit{ Phys. Rev. C} \textbf{ 97} (2018) 024904,
  \href{http://dx.doi.org/10.1103/PhysRevC.97.024904}{\doi{10.1103/PhysRevC.97.024904}},
\href{http://www.arXiv.org/abs/1708.03559}{\texttt{arXiv:1708.03559}}.

\bibitem{Huo:2017nms}
\hrefCMSnoop {}{P.~Huo, K.~Gajdosova, J.~Jia, and Y.~Zhou, ``{Importance of
  non-flow in mixed-harmonic multi-particle correlations in small collision
  systems}'',} \textit{ Phys. Lett. B} \textbf{ 777} (2018) 201,
  \href{http://dx.doi.org/10.1016/j.physletb.2017.12.035}{\doi{10.1016/j.physletb.2017.12.035}},
\href{http://www.arXiv.org/abs/1710.07567}{\texttt{arXiv:1710.07567}}.

\bibitem{Chatrchyan:2014fea}
\hrefCMSnoop {}{{CMS Collaboration}, ``{Description and performance of track
  and primary-vertex reconstruction with the CMS tracker}'',} \textit{ JINST}
  \textbf{ 9} (2014) P10009,
  \href{http://dx.doi.org/10.1088/1748-0221/9/10/P10009}{\doi{10.1088/1748-0221/9/10/P10009}},
\href{http://www.arXiv.org/abs/1405.6569}{\texttt{arXiv:1405.6569}}.

\bibitem{GEANT4}
\hrefCMSnoop {}{{Geant4} Collaboration, ``{Geant4} --- a simulation toolkit'',}
  \textit{ Nucl. Instrum. Meth. A} \textbf{ 506} (2003) 250,
\href{http://dx.doi.org/10.1016/S0168-9002(03)01368-8}{\doi{10.1016/S0168-9002(03)01368-8}}.

\bibitem{Chatrchyan:2008zzk}
\hrefCMSnoop {}{{CMS Collaboration}, ``The {CMS} experiment at the {CERN}
  {LHC}'',} \textit{ JINST} \textbf{ 3} (2008) S08004,
\href{http://dx.doi.org/10.1088/1748-0221/3/08/S08004}{\doi{10.1088/1748-0221/3/08/S08004}}.

\bibitem{PhysRevAccelBeams.20.081003}
\hrefCMSnoop {}{E.~Todesco and J.~Wenninger, ``Large hadron collider momentum
  calibration and accuracy'',} \textit{ Phys. Rev. Accel. Beams} \textbf{ 20}
  (Aug, 2017) 081003,
  \href{http://dx.doi.org/10.1103/PhysRevAccelBeams.20.081003}{\doi{10.1103/PhysRevAccelBeams.20.081003}}.

\bibitem{Chatrchyan:2013nka}
\hrefCMSnoop {}{{CMS} Collaboration, ``{Multiplicity and transverse momentum
  dependence of two- and four-particle correlations in \pPb\ and \PbPb\
  collisions}'',} \textit{ Phys. Lett. B} \textbf{ 724} (2013) 213,
  \href{http://dx.doi.org/10.1016/j.physletb.2013.06.028}{\doi{10.1016/j.physletb.2013.06.028}},
\href{http://www.arXiv.org/abs/1305.0609}{\texttt{arXiv:1305.0609}}.

\bibitem{Khachatryan:2016bia}
\hrefCMSnoop {}{{CMS Collaboration}, ``{The CMS trigger system}'',} \textit{
  JINST} \textbf{ 12} (2017) 01020,
  \href{http://dx.doi.org/10.1088/1748-0221/12/01/P01020}{\doi{10.1088/1748-0221/12/01/P01020}},
\href{http://www.arXiv.org/abs/1609.02366}{\texttt{arXiv:1609.02366}}.

\bibitem{Khachatryan:2016got}
\hrefCMSnoop {}{{CMS Collaboration}, ``{Observation of charge-dependent
  azimuthal correlations in $p$-Pb collisions and its implication for the
  search for the chiral magnetic effect}'',} \textit{ Phys. Rev. Lett.}
  \textbf{ 118} (2017) 122301,
  \href{http://dx.doi.org/10.1103/PhysRevLett.118.122301}{\doi{10.1103/PhysRevLett.118.122301}},
\href{http://www.arXiv.org/abs/1610.00263}{\texttt{arXiv:1610.00263}}.

\bibitem{Sirunyan:2017quh}
\hrefCMSnoop {}{{CMS Collaboration}, ``{Constraints on the chiral magnetic
  effect using charge-dependent azimuthal correlations in $p\mathrm{Pb}$ and
  PbPb collisions at the CERN Large Hadron Collider}'',} \textit{ Phys. Rev. C}
  \textbf{ 97} (2018) 044912,
  \href{http://dx.doi.org/10.1103/PhysRevC.97.044912}{\doi{10.1103/PhysRevC.97.044912}},
  \href{http://www.arXiv.org/abs/1708.01602}{\texttt{arXiv:1708.01602}}.

\bibitem{Ollitrault:2009ie}
\hrefCMSnoop {}{J.-Y. Ollitrault, A.~M. Poskanzer, and S.~A. Voloshin,
  ``{Effect of flow fluctuations and nonflow on elliptic flow methods}'',}
  \textit{ Phys. Rev. C} \textbf{ 80} (2009) 014904,
  \href{http://dx.doi.org/10.1103/PhysRevC.80.014904}{\doi{10.1103/PhysRevC.80.014904}},
\href{http://www.arXiv.org/abs/0904.2315}{\texttt{arXiv:0904.2315}}.

\bibitem{Khachatryan:2015oea}
\hrefCMSnoop {}{{CMS Collaboration}, ``{Evidence for transverse momentum and
  pseudorapidity dependent event plane fluctuations in PbPb and pPb
  collisions}'',} \textit{ Phys. Rev. C} \textbf{ 92} (2015) 034911,
  \href{http://dx.doi.org/10.1103/PhysRevC.92.034911}{\doi{10.1103/PhysRevC.92.034911}},
\href{http://www.arXiv.org/abs/1503.01692}{\texttt{arXiv:1503.01692}}.

\end{thebibliography}\endgroup
